\documentclass[12pt]{article}

\usepackage{amsmath,amssymb,amscd,latexsym,anysize,setspace}

\usepackage{times,mathptm,color}

\marginsize{1in}{1in}{1in}{1in}

\newcommand{\ket}[1]{|#1\rangle}
\newcommand{\bra}[1]{\langle #1 |}
\newcommand{\bracket}[2]{\langle #1 | #2 \rangle}

\newcommand{\CC}{\mathbf{C}}
\newcommand{\RR}{\mathbf{R}}

\newcommand{\ent}{\varepsilon}

\newcommand{\notGate}{
    \put(0,1){\line(1,0){2}}
    \put(1,1){\circle{1}}
    \put(1,0.5){\line(0,1){1}}
}

\newcommand{\botCNOT}{
\begin{picture}(0,0)
    \put(1,3){\circle{1}}
    \put(0,1){\line(1,0){2.3}}
    \put(1,1){\line(0,1){2.5}}
    \put(1,1){\circle*{0.4}}
    \put(0,3){\line(1,0){2.3}}
\end{picture}
}

\newcommand{\topCNOT}{
\begin{picture}(0,0)
    \put(1,3){\circle*{0.4}}
    \put(0,3){\line(1,0){2.3}}

    \put(1,1){\circle{1}}
    \put(1,0.5){\line(0,1){2.5}}
    \put(0,1){\line(1,0){2.3}}
\end{picture}
}

\newcommand{\topC}{
\begin{picture}(0,0)
    \put(1,3){\circle*{0.4}}
    \put(1,3){\line(0,-1){1.3}}
    \put(0,3){\line(1,0){2}}
\end{picture}
}

\newcommand{\boxGate}[1]{
\begin{picture}(0,0)
    \put(0.3,0.3){\framebox(1.4,1.4){\small #1}}
    \put(0,1){\line(1,0){0.3}}
    \put(1.7,1){\line(1,0){0.3}}
\end{picture}
}

\newcommand{\boxGateTwo}[1]{
\begin{picture}(0,0)
    \put(0.3,0.3){\framebox(1.4,3.4){\small #1}}
    \put(0,1){\line(1,0){0.3}}
    \put(1.7,1){\line(1,0){0.3}}
    \put(0,3){\line(1,0){0.3}}
    \put(1.7,3){\line(1,0){0.3}}
\end{picture}
}

\newcommand{\hWire}{
   \begin{picture}(0,0)
    \put(0,1){\line(1,0){2}}
   \end{picture}
}

\newtheorem{theorem}{Theorem}
\newtheorem{proposition}[theorem]{Proposition}

\begin{document}

\newcommand{\proof}[1]{{\bf Proof:} #1 $\blacksquare$}

\title{Quantum Circuits for Incompletely Specified  \\Two-Qubit
Operators}

\author{Vivek V. Shende \and Igor L. Markov \and \small Department of Electrical Engineering
and Computer Science \\ \small The University of Michigan, Ann
Arbor, Michigan, 48109-2212, USA}

\maketitle

\abstract{While the question ``how many {\tt CNOT} gates are
needed to simulate an arbitrary two-qubit operator'' has been
conclusively answered -- three are necessary and sufficient --
previous work on this topic assumes that one wants to simulate a
given unitary operator up to global phase. However, in many
practical cases additional degrees of freedom are allowed. For
example, if the computation is to be followed by a given
projective measurement, many dissimilar operators achieve the same
output distributions on all input states. Alternatively, if it is
known that the input state is $\ket{0}$, the action of the given
operator on all orthogonal states is immaterial. In such cases, we
say that the unitary operator is incompletely specified; in this
work, we take up the practical challenge of satisfying a given
specification with the smallest possible circuit. In particular,
we identify cases in which such operators can be implemented using
fewer quantum gates than are required for generic completely
specified operators.}

\doublespace

\section{Introduction}

  Quantum circuits offer a common formalism to describe
  various quantum-mechanical effects and facilitate
  a unified framework for simulating such effects on a quantum
  computer \cite{Nielsen:00}.  The framework consists of two steps:
  (1) for a given unitary evolution, find a quantum circuit
  that computes it,  (2) implement this circuit on a quantum computer.
  The first step is sometimes called quantum circuit synthesis
  \cite{Bullock:03}, and is the focus of our work.
  Given that existing physical implementations are severely limited
  by the number of qubits, a considerable effort was made recently
  to synthesize small two-qubit circuits \cite{Vidal:04,
  Vatan:04,Shende:18gates:04,Shende:cnotcount:04,
  Zhang:Bgate:04}. It has been shown that
  for such a circuit to implement a typical two-qubit
  operator, three {\tt CNOT} gates are needed. However,
  this result was proven under the assumption that we know nothing
  about the circuit surrounding the given two-qubit operator.
  Thus, in the event that we have additional information, say the
  state of the input qubits or the basis in which the result of the
  computation is to be measured, the result no longer holds.
  In fact, we show that if the input state is $\ket{0}$, then
  three one-qubit gates and one {\tt CNOT} suffice to simulate
  an arbitrary two-qubit operator. We also show that if a projective
  measurement in the computational basis follows the two-qubit
  operator, then it can be implemented by a circuit with
  two {\tt CNOT}s.


\section{Background} \label{sec:background}

The following family of ``spin flip'' or ``$\sigma_y \otimes
\sigma_y$'' results are invaluable in the study of two-qubit
operators. They are all related in some sense to the fact that a
two-qubit pure state $\ket{\phi}$ is separable if and only if
$\ent(\ket{\phi}) := \bra{\phi^*} \sigma_y^{\otimes 2} \ket{\phi}
= 0$. For this reason, $|\ent|^2$ is sometimes used to measure
entanglement.\\

\noindent {\bf Facts about two-qubit operators.}

\begin{enumerate}
\item {\bf The Magic Basis} \cite{Bennett:96, Hill:97}. There
exist matrices $E \in U(4)$ such that $E^\dag SO(4) E =
SU(2)^{\otimes 2}$. These are characterized by the property $EE^T
= \sigma_y^{\otimes 2}$.

\item {\bf The Makhlin Invariants \cite{Makhlin:02}} Let $u, v \in
SU(4)$. Then there exist $a,b,c,d \in SU(2)$ such that $(a \otimes
b) u (c \otimes d) = v$ if and only if $u^T \sigma_y^{\otimes 2}u
\sigma_y^{\otimes 2}$ and $v^T \sigma_y^{\otimes 2}v
\sigma_y^{\otimes 2}$ have the same spectrum.

\item {\bf The Canonical Decomposition
\cite{Khaneja:01,Lewenstein:01}} Any $u \in SU(4)$ can be written
in the following form. \[u = (a \otimes b) e^{i(
I \otimes I + \theta_x \sigma_x \otimes \sigma_x + \theta_y \sigma_y \otimes
\sigma_y + \theta_z \sigma_z \otimes \sigma_z)} (c \otimes d)\]
Above, $a,b,c,d \in SU(2)$ and $\theta_x, \theta_y, \theta_z \in
\RR$.
\end{enumerate}

  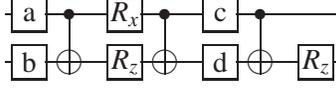
\begin{figure}
    \begin{center}
    \begin{picture}(14,4)
        \put(0,0){\boxGate{b}}
        \put(0,2){\boxGate{a}}
        \put(2,0){\topCNOT}
        \put(4,0){\boxGate{$R_z$}}
        \put(4,2){\boxGate{$R_x$}}
        \put(6,0){\topCNOT}
        \put(8,0){\boxGate{d}}
        \put(8,2){\boxGate{c}}
        \put(10,0){\topCNOT}
        \put(12,0){\boxGate{$R_z$}}
        \put(12,2){\hWire}
    \end{picture}
    \caption{\label{fig:18:cxz} A universal two-qubit
    circuit with three {\tt CNOT} gates \cite{Shende:18gates:04}.
    It contains seven one-qubit placeholders, which can be translated
    into 15 placeholders for one-parameter gates.}
    \end{center}
  \end{figure}

These facts can be used to classify two-qubit pure states up to
the action of local unitaries, as shown below, and this result is
used later in our work. One can also classify mixed states, but
the more general result is harder to state \cite{Makhlin:02}.

\begin{proposition} \label{prop:2qfacts} Let $\ket{\phi}$ and
$\ket{\psi}$ be 2-qubit pure states. Then $\ket{\phi}$ and
$\ket{\psi}$ can be interchanged by local unitary operators if and
only if $|\ent(\ket{\phi})| = |\ent(\ket{\psi})|$.
\end{proposition}
\proof{($\Rightarrow$) Suppose first that $\phi$ and $\psi$ are
interconvertible by local unitaries, that is, there exist $a, b
\in U(2)$ such that $(a \otimes b) \ket{\phi} = \ket{\psi}$. One
can check that for $m$ a $2 \times 2$ matrix, $m^T \sigma_y m =
\sigma_y \det m$. Using this to simplify, we have $\bra{\psi^*}
\sigma_y^{\otimes 2} \ket{\psi} = \bra{\phi^*} (a \otimes b)^T
\sigma_y^{\otimes 2} (a \otimes b) \ket{\phi} = (\det a)(\det b)
\bra{\phi^*} \sigma_y \ket{\phi}$. The scalar vanishes upon taking
absolute value. ($\Leftarrow$) Conversely, suppose $|\bra{\phi^*}
\sigma_y \otimes \sigma_y \ket{\phi}| = |\bra{\psi^*} \sigma_y
\otimes \sigma_y \ket{\psi}|$. By ignoring global phase, we may
suppose that in fact $\bra{\phi^*} \sigma_y \otimes \sigma_y
\ket{\phi}= \bra{\psi^*} \sigma_y \otimes \sigma_y \ket{\psi}$.
Changing to the Magic Basis transforms the hypothesis into
$\langle \phi | \phi \rangle = \langle \psi | \psi \rangle$, and
the statement we want to prove into: there exists $p \in SO(4)$
such that $p \ket{\psi} = \ket{\phi}$. So, let $v \in \CC^4$ be an
arbitrary vector, and $v = v_r + iv_i$ be its decomposition into
real and imaginary parts. Then we see that $v^T v = |v_r|^2 -
|v_i|^2 + 2i v_r^T v_i$. Since we know $v$ to be a unit vector,
$v^T v$ encodes the magnitudes of the real and imaginary parts of
$v$, and the angle between them. From this it is clear that two
unit vectors $v,w$ in $\CC^4$ can be interchanged by an element of
$SO(4)$ if and only if $v^T v = w^T w$, and we have proven our
claim.}

Another important consequence of the $\sigma_y \otimes \sigma_y$
theorems is the result that an arbitrary two-qubit operator can be
implemented by a circuit containing three {\tt CNOT}s and some
one-qubit gates. It has been proven in various forms
\cite{Vidal:04,Vatan:04,Shende:18gates:04}, of which we need the
particular one described in Fig. \ref{fig:18:cxz}.

It is also known that three {\tt CNOT} gates are necessary to
implement some two-qubit operators, such as a wire swap
\cite{Vidal:04,Vatan:04,Shende:18gates:04}. To prove this and
other lower bound results, one considers {\em generic circuits}.
These are diagrams with placeholders for unspecified (variable)
gates and may also contain specific (constant) gates. Each
placeholder corresponds to some subset of possible gates. In this
work, all placeholders are for one-qubit gates, and all constant
gates are either {\tt CNOT}s or one-qubit gates; we call such
circuits {\em basic}. We label a placeholder for an unspecified
element of $SU(2)$ with a lower-case roman letter, and
placeholders for gates of the form $R_x(\alpha)$, $R_y(\beta)$, or
$R_z(\gamma)$ by $R_x, R_y, R_z$ respectively. Here, $R_n(\theta)
= e^{i \sigma^n \theta/2}$. We refer to basic circuits whose only
placeholders represent $R_x, R_y, R_z$ gates as {\em elementary}.
The motivation for restricting to elementary circuits is that each
placeholder has one degree of freedom, which makes dimension
counting easier and more precise. Moreover, nothing is lost by
doing so since any $u \in SU(2)$ can be written in the form
$R_k(\alpha) R_l(\beta) R_m(\gamma)$ for any $k,l,m \in \{x,y,z\},
k\ne l, l \ne m$.

We say that an $n$-qubit generic circuit is {\em universal} if, by
specifying appropriate values for the placeholders, one can obtain
a circuit simulating arbitrary $u \in U(2^n)$ up to global phase.
The dimension of $U(2^n)$ is $4^n$; subtracting one for global
phase, we see that an elementary circuit on $n$ qubits cannot be
universal unless there are at least $4^n - 1$ placeholders. Our
general strategy for showing that a given incompletely specified
circuit is not universal is to convert it into an elementary
circuit, eliminate as many placeholders as possible via circuit
identities, and then count gates. For example, the following
well-known identity is particularly instrumental:
the $R_x$ (respectively, $R_z$) gate can pass through
the target (respectively, control) of a {\tt CNOT} gate.

\section{Preparation of Pure States}\label{sec:prep}

  The three-{\tt CNOT} lower bound applies when one must find
  a circuit to simulate a particular given two-qubit operator up
  to global phase. However, quantum-computational tasks arising
  in applications are often less completely specified, thus they
  can be performed by a greater variety of quantum circuits.
  One such task is state preparation. To prepare the $n$-qubit
  state $\ket{\phi}$ from $\ket{0}$, we can use any operator $u \in
  U(2^n)$ with $u\ket{0} = e^{i \theta} \ket{\phi}$.
  A poor choice of $u$
  ensures that $u$ cannot be implemented with fewer than $O(4^n)$
  gates. However, as the dimension of the space of pure states is
  $2^n - 1$, the lower bound by dimension counting techniques described
  in Section \ref{sec:background}
  only indicate that at least $\lceil(2^n - 3n -1)/4\rceil$ {\tt CNOT}s are
  necessary to prepare an arbitrary pure state.\footnote{For more
  details on the use of these
  lower bounds methods, see Section 3 of
  \cite{Shende:18gates:04}.}\;
  We show below that
  this bound can be matched asymptotically by techniques based on
   the QR decomposition of matrices.

  \begin{proposition} \label{prop:prepn}
      Preparing a generic $n$-qubit pure state from $\ket{0}$
      requires $O(2^n)$ gates.
  \end{proposition}
  \proof{As shown in \cite{Vartiainen:04}, an arbitrary $n$-qubit
  unitary operator can be simulated by a circuit containing
  approximately $8.7 \times 4^n$ {\tt CNOT} gates. Their
  technique is based on the QR decomposition and gives a circuit that
  builds up a unitary matrix column by column, with each of the $2^n$ columns
  built by a subcircuit containing $O(2^n)$ gates. For our present
  purposes, only the subcircuit responsible for the first
  column is needed.}

   Other decomposition algorithms find better circuits for arbitrary
   operators: the best currently known yields about $4^n / 2$
   {\tt CNOT}s \cite{Shende:NQ:04}
   and is a factor of two away from the
   lower bound of $\lceil(4^n - 3n - 1)/4\rceil$ given in
   \cite{Shende:18gates:04}. However, as these algorithms
   do not build matrices column by
   column, they do not yield efficient techniques for
   state preparation. We note in passing that a significantly
   larger gap exists between the upper and lower bounds on the
   number of {\tt CNOT} gates needed to prepare an arbitrary
   state, as compared to the corresponding bounds for the problem
   of simulating an arbitrary unitary operator: in the first case,
   a factor of thirty, in the second, a factor of two.\footnote{Since
   the first posting of this paper,
   several preprints have appeared to address this gap. In particular,
   it has been shown in \cite{Shende:NQ:04,Mottonen:04}
   that $2^{n+1} - 2n - 2$ {\tt CNOT} gates suffice to
   prepare an arbitrary $n$-qubit state from $\ket{0}$.
   A different technique based on Grover's Search Algorithm also purports
   to do well in some special cases \cite{Soklakov:04}.}

  We now seek optimality results for the task of state preparation
  in the case of two qubits. As two-qubit states can be entangled,
  at least one use of a two-qubit gate is necessary to prepare
  any entangled state. To characterize two-qubit gates which are also
  sufficient for this purpose, we use some
  concepts from algebraic geometry, for whose
  explication the reader is referred to any introductory textbook,
  such as \cite{Shafarevich}. We also give an
  explicitly constructive proof of this result in the special case
  of the {\tt CNOT} gate.

  \begin{proposition} \label{prop:prep2} Let $G \in SU(4)$.
  Then an arbitrary pure state $\ket{\psi}$ can be prepared
  from $\ket{0}$ by a circuit containing one-qubit gates and
  a single gate $G$ if and only if there exists a state $\ket{\phi}$
  such that $\ent(\ket{\phi}) = 0$ and $\ent(G \ket{\phi}) = 1$.
  \end{proposition}
  \proof{
  ($\Leftarrow$) Note that $|\ent(\ket{0})| = 0$.
  Define $\ket{B} := (\ket{00} + \ket{11})/\sqrt{2}$
  so $|\ent(\ket{B})| = 1$. Suppose
  there exist $a,b,c,d \in U(2)$ such that $(a \otimes b) G (c \otimes d)
  \ket{0} = \ket{B}$. Recalling from Proposition \ref{prop:2qfacts} that
  one-qubit operators preserve $|\ent|$, we have
  $|\ent((c \otimes d) \ket{0})| = 0$, and
  $|\ent(G(c \otimes d) \ket{0})| = |\ent{(\ket{B})}| = 1$.

  ($\Rightarrow$) We note that by Proposition
  \ref{prop:2qfacts}, it suffices to show that circuits of the
  form $G (c \otimes d)$ can prepare states with
  arbitrary $|\ent|$ from $\ket{0}$. Again by Proposition
  \ref{prop:2qfacts}, if $\ket{\phi}$ is the state given in the
  hypothesis, then there exist $a_1,b_1 \in U(2)$ such that $(a_1
  \otimes b_1) \ket{0} = \ket{\phi}$. So, $\ent( G
  (a_1 \otimes b_1) \ket{0}) = 1$. If we show that a state
  with $|\ent| = 0$ can be prepared, it will follow by continuity
  of $|\ent|$ that arbitrary states can be prepared as well.

  It suffices to show that every
  two-qubit gate maps some $|\ent| = 0$ state to
  another. For, if $\ket{\phi}$ is such a state
  for $G$, then we may choose $a_0,b_0 \in U(2)$ such that
  $\ket{\phi} = (a_0 \otimes b_0)\ket{0}$, and see that
  $|\ent(G(a \otimes b) \ket{0})| = 0$. Thus it
  suffices show that $\ent(\ket{\phi}) = 0$ and
  $\ent(G(\ket{\phi})) = 0$ have common solutions for all $G$.
  Fix $G$, and fix a basis for the state space. Then,
  $\ent(\ket{\phi})$ and $\ent(G \ket{\phi})$
  can be seen to be homogenous polynomials in the
  4 coordinates (in fact, they are quadratic forms).
  In particular, the zeroes of these polynomials do not depend
  on global phase, so we may speak of their zeroes on the
  space $\mathbf{CP}^3$ of two-qubit pure states modulo global
  phase. It is a fact that any two (nonconstant) homogenous polynomials
  must have common zeroes here \cite{Shafarevich}.
}

  As a single {\tt CNOT} and a Hadamard gate can be used to
  prepare $(\ket{00} + \ket{11})/\sqrt{2})$ from $\ket{00}$, the
  {\tt CNOT} gate satisfies the hypothesis of the
  Proposition \ref{prop:prep2}, and therefore a single {\tt CNOT} suffices to
  prepare an arbitrary two-qubit pure state from
  $\ket{0}$. We now give a more explicit construction in this
  case.

\begin{proposition} \label{prop:prep2c}
  A two-qubit pure state $\ket{\phi}$ can be prepared from $\ket{0}$ using the
  one {\tt CNOT} gate and three one-qubit gates.
\end{proposition}
\proof{
  Let $C_2^1$ be the {\tt CNOT} gate controlled on the higher qubit and
  acting on the lower. Let $c = u
  \ket{0}\bra{0} + v \ket{1}\bra{0} - \overline{v} \ket{0}\bra{1}
  + \overline{u} \ket{1}\bra{1}$ for some $u, v \in \CC$; one can
  check that $c \in SU(2)$. Let $\phi_i = \bracket{i}{\phi}$.
  We explicitly compute
    \[\ent(C_2^1(I \otimes c)\ket{\phi}) =
  \phi_0 \phi_2 (u^2 - v^2) +
  \phi_1 \phi_3 (\overline{v}^2 - \overline{u}^2) -
  (\phi_0 \phi_3 + \phi_1 \phi_2)(u\overline{v} + v
  \overline{u})\]
  Making the change of variables $z = u^2 - v^2$, $\lambda = (u\overline{v} + v
  \overline{u})$, we note that $\lambda \in \RR$ and $|z|^2 + \lambda^2 = 1$;  we
  want to solve $\phi_0 \phi_2 z - \phi_1 \phi_3 \overline{z} =
  (\phi_0 \phi_3+\phi_1 \phi_2) \lambda$ for $z, \lambda$. This is a linear
  system with two equations and three unknowns; thus we obtain
  $z, \lambda$ up to a scalar multiple,
  and can choose the scalar so that $|z|^2 + \lambda = 1$.

  Let $\ket{\eta} = C_2^1(I \otimes c)\ket{\phi}$ and
  verify that $\ent(\ket{\eta}) = 0$. Since $\ket{\eta}$ is separable,
  write it as $\ket{s} \ket{t}$. This allows one to define
  $a$ and $b$ so that $(a\otimes b)\ket{0}=\ket{s} \ket{t}$.
  Finally, we can write
  $(I \otimes c^\dag)C_2^1(a \otimes b) \ket{0} = \ket{\phi}$ as
  desired.
}

\section{Measurement Don't-Cares} \label{sec:meas}

  Fewer gates are required for state preparation because
  images of basis states other than $\ket{0}$ can be arbitrary
  (in other words, we are using additional information about the input).
  Similarly, we may be able to save gates if we know in
  advance how the circuit output will be used. In particular,
  we now suppose that we know the output is to be measured
  in some predetermined basis.

  Suppose we intend to first simulate an
  operator $u$ on a yet-unspecified input, then take a
  projective measurement with respect to some given orthogonal
  subspace decomposition $(\CC^2)^{\otimes n} = \bigoplus E_i$,
  and we are interested only in having the measured state appear
  in a given subspace with the appropriate probability. In
  particular, if $v$ is an operator that preserves each subspace
  $E_i$, then we do not care whether we implement $u$ or $vu$.
  Conversely, if $w$ is any operator which, upon any input, agrees
  with $u$ after projective measurement with respect to the given
  subspace decomposition, then it is clear that $wu^{-1}$
  preserves each subspace $E_i$. If a given circuit simulates
  some such operator $w$ up to phase, we say that this circuit simulates
  $u$ up to the measurement don't care associated to the given
  subspace decomposition.

  Mathematically speaking, the problem of state preparation is
  essentially a special case of a measurement don't care.
  To prepare the state $\ket{\phi}$ from $\ket{0}$, it is enough
  to have any operator whose matrix in the computational basis has
  first column $\ket{\phi}$. On the other hand, suppose we are
  interested in simulating some given operator $u$, then
  taking a projective measurement with respect to two orthogonal
  subspaces: one spanned by $\ket{0}$ and the other by the rest of
  the computational basis vectors. Then we may replace $u$ with
  any operator $v$ such that $\bra{0}u = \bra{0}v$; that is, the matrices
  of $u$ and $v$ must have the same first row in the computational
  basis. Thus the problem of state preparation amounts to
  specifying a single column of a matrix, whereas the
  aforementioned measurement don't care amounts to specifying a
  single row. Thus Propositions \ref{prop:prepn},
  \ref{prop:prep2}, and \ref{prop:prep2c} carry over to this
  context.

  \begin{proposition} \label{prop:meas1dim}
    To simulate an arbitrary $n$-qubit operator up to a projective
    measurement onto two subspaces, one of which is one
    dimensional, at least $\lceil(2^n - 3n - 1)/4\rceil$ {\tt CNOT} gates
    are necessary, and $O(2^n)$ {\tt CNOT} gates are sufficient.
    For $n=2$, one {\tt CNOT} is necessary and sufficient.
  \end{proposition}

  Suppose now we have a subspace decomposition and an
  underspecified circuit $S$ which we believe is universal up to
  the associated measurement don't care -- that is, we believe
  that for any $u$, appropriate specification of parameters gives
  a circuit simulating an operator $w$ such that there exists some
  operator $v$ preserving the subspace decomposition with $vu = w$.
  Let $T$ be an underspecified circuit that precisely captures the set
  of operators that fix the subspace decomposition.
  It is clear that $S$ is universal up to the given measurement don't care
  if and only if the concatenated circuit $ST$ is universal. Therefore,
  as we show below, one cannot claim asymptotic savings for this
  problem in general.

  \begin{proposition}
    To simulate an arbitrary $n$-qubit operator up to a projective
    measurement in which each of the subspaces is one-dimensional,
    at least $\lceil(4^n - 2^n - 3n)/4\rceil$ {\tt CNOT} gates are required.
  \end{proposition}
  \proof{
    First, note that an operator $v$ can be right-multiplied by any
    diagonal operator $\delta$ (diagonal in the basis of the
    measurement) and that the group of diagonal matrices is
    $2^n$-dimensional. $4^n - 2^n$ parameters remain to be
    accounted for, and the proof of Proposition 1 of
    \cite{Shende:18gates:04} indicates that $(4^n - 2^n - 3n)/4$ {\tt CNOT} gates
    are necessary to account for this many parameters.
  }

  Given that the best known circuit synthesis technique for
  $n$-qubit circuits is still a factor of two away from the
  theoretical lower bound of $\lceil(4^n - 3n -1) / 4\rceil$, it may be difficult
  to detect a savings of $2^n$ gates by analyzing specific circuits.
  Thus we turn to the two-qubit case, where all bounds are known,
  tight, and small --- no more than three {\tt CNOT} gates are required,
  and a savings of even one gate would be significant.

  On two qubits, there are several different types of measurement possible.
  We classify them by the subspace dimensions, hence we have ``$3+1$'', ``$2+2$'',
  ``$2+1+1$'', and ``$1+1+1+1$'' measurements. In what
  follows, we generally require that each subspace is spanned by computational
  basis vectors. We refer collectively to the corresponding measurement
  don't-cares as CB-measurement don't-cares.
  Additionally, when dealing with $2+2$ measurements, we assume that
  one of the qubits is measured; that is, we do not consider the
  decomposition $\CC^4 = \mbox{span}(\ket{00},\ket{11}) \oplus
  \mbox{span}(\ket{01}, \ket{10})$. Indeed, measuring a qubit is a
  common step in quantum algorithms and communication protocols.

  We have already seen in Proposition \ref{prop:meas1dim}
  that one {\tt CNOT} is necessary and sufficient in
  the $3+1$ case. We now show that at
  least two {\tt CNOT}s are needed in the remaining cases

  \begin{proposition} \label{prop:meas2qlbounds}
    Let $\CC^4 = \bigoplus E_i$ be a CB-subspace decomposition corresponding to the
    measurement don't-care $M$. Suppose no subspace is
    $3$-dimensional, and further that the subspace decomposition is
    not $\CC^4 = \{\mathrm{span}(\ket{00},\ket{11}) \oplus
    \mathrm{span}(\ket{01},\ket{10})$.\footnote{
  In the $2+2$ case where measurement is performed
  ``across qubits'', the key question is whether \noindent
  \begin{center}
    \begin{picture}(14,4)
      \put(0,0){\boxGate{b}}
      \put(0,2){\boxGate{a}}
      \put(2,0){\topCNOT}
      \put(4,0){\boxGate{d}}
      \put(4,2){\boxGate{c}}
      \put(6,0){\botCNOT}
      \put(8,0){\boxGate{e}}
      \put(8,2){\boxGate{$R_z$}}
      \put(10,0){\boxGate{f}}
      \put(10,0){\topC}
      \put(12,0){\botCNOT}
    \end{picture}
  \end{center}
  is universal. Unfortunately, we have neither been able to find
  circuit identities to reduce the number of one-parameter gates
  below $15$, nor to show that this circuit is universal.}
    Then there exist two-qubit operators that cannot be simulated up to $M$ by
    a circuit with only one {\tt CNOT}.
  \end{proposition}
  \proof{
    First, consider subspace decompositions in which neither
    $\mathrm{span}(\ket{00},\ket{11})$,
    nor $\mathrm{span}(\ket{01},\ket{10})$ occur.
    Remaining cases with 2+1+1 decompositions using one of those subspaces
    are considered separately below.
    Suppose an operator is universal up to a $1+1+1+1$ or $2+1+1$
    CB-measurement
    don't-care satisfying the above condition. Then combining pairs of
    $1$-dimensional subspaces into $2$-dimensional subspaces,
    we see that the same circuit is universal up to a $2+2$
    CB-measurement don't-care in which one of the qubits is
    measured. Suppose, without loss of generality, that it is
    the higher order qubit, hence that the subspaces are
    $\mathrm{span}(\ket{00},\ket{01})$ and
    $\mathrm{span}(\ket{10},\ket{11})$.

    We now compose an arbitrary one-{\tt CNOT} circuit with a circuit for
    operators preserving the relevant CB-subspaces, as outlined
    at the beginning of the section (see below-left).
    Conglomerating adjacent gates, we obtain the circuit
    below-right.
    
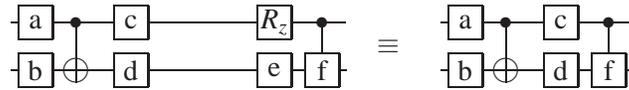
\begin{center}
        \begin{picture}(26,4)
            \put(0,0){\boxGate{b}}
            \put(0,2){\boxGate{a}}
            \put(2,0){\topCNOT}
            \put(4,0){\boxGate{d}}
            \put(4,2){\boxGate{c}}
            \put(6,0){\hWire}
            \put(6,2){\hWire}
            \put(8,0){\hWire}
            \put(8,2){\hWire}
            \put(10,0){\boxGate{e}}
            \put(10,2){\boxGate{$R_z$}}
            \put(12,0){\boxGate{f}}
            \put(12,0){\topC}

            \put(15.6,1.7){$\equiv$}

            \put(18,0){\boxGate{b}}
            \put(18,2){\boxGate{a}}
            \put(20,0){\topCNOT}
            \put(22,0){\boxGate{d}}
            \put(22,2){\boxGate{c}}
            \put(24,0){\boxGate{f}}
            \put(24,0){\topC}
        \end{picture}
    \end{center}
    We now convert 3-dimensional place-holders to one-parameter gates,
    pass $R_x$ and $R_z$ through {\tt CNOT} where desirable, and
    conglomerate adjacent gates again.
    
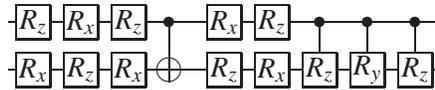
\begin{center}
        \begin{picture}(18,4)
            \put(0,0){\boxGate{$R_x$}}
            \put(0,2){\boxGate{$R_z$}}
            \put(2,0){\boxGate{$R_z$}}
            \put(2,2){\boxGate{$R_x$}}
            \put(4,0){\boxGate{$R_x$}}
            \put(4,2){\boxGate{$R_z$}}
            \put(6,0){\topCNOT}
            \put(8,0){\boxGate{$R_z$}}
            \put(8,2){\boxGate{$R_x$}}
            \put(10,0){\boxGate{$R_x$}}
            \put(10,2){\boxGate{$R_z$}}
            \put(12,0){\boxGate{$R_z$}}
            \put(12,0){\topC}
            \put(14,0){\boxGate{$R_y$}}
            \put(14,0){\topC}
            \put(16,0){\boxGate{$R_z$}}
            \put(16,0){\topC}
        \end{picture}
    \end{center}
    As this circuit has 13 one-parameter gates,
    the circuit we started with cannot be universal.

    The $2+1+1$ cases where the $2$-dimensional subspace is
    $\mathrm{span}(\ket{00},\ket{11})$ or
    $\mathrm{span}(\ket{01},\ket{10})$ can be dealt with similarly. We
    give the circuits preserving these subspace decompositions below;
    the left circuit corresponds to $\mathrm{span}(\ket{01},\ket{10})$
    and the right to $\mathrm{span}(\ket{01},\ket{10})$.
    \begin{center}
        \begin{picture}(24,4)
            \put(0,0){\botCNOT}
            \put(2,0){\boxGate{$R_z'$}}
            \put(2,2){\boxGate{$R_z$}}
            \put(4,0){\topC}
            \put(4,0){\boxGate{e}}
            \put(6,0){\botCNOT}

            \put(12,2){\hWire}
            \put(12,0){\notGate}
            \put(14,0){\botCNOT}
            \put(16,0){\boxGate{$R_z'$}}
            \put(16,2){\boxGate{$R_z$}}
            \put(18,0){\topC}
            \put(18,0){\boxGate{e}}
            \put(20,0){\botCNOT}
            \put(22,0){\notGate}
            \put(22,2){\hWire}
        \end{picture}
    \end{center}
    In both cases, the placeholder marked $R_z'$ can be conglomerated
    with another placeholder, leaving a circuit with 14 one-parameter
    placeholders.
  }

  It is a natural question whether one might do better with a
  different gate \cite{Zhang:Bgate:04}. At least for the $1+1+1+1$
  subspace decomposition, the answer is no.

  \begin{proposition} \label{prop:meas2qlbounds:anygate}
    Fix a two-qubit gate $G$.
    Some two-qubit operators cannot be simulated, up to
    the $1+1+1+1$ CB-measurement don't care, by a circuit
    with a single instance of $G$.
  \end{proposition}
  \proof{
  Compose the circuit in question with a circuit for
  simulating a diagonal operator.
    
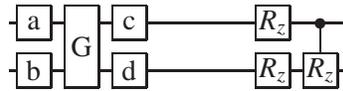
\begin{center}
        \begin{picture}(14,4)
            \put(0,0){\boxGate{b}}
            \put(0,2){\boxGate{a}}
            \put(2,0){\boxGateTwo{G}}
            \put(4,0){\boxGate{d}}
            \put(4,2){\boxGate{c}}
            \put(6,0){\hWire}
            \put(6,2){\hWire}
            \put(8,0){\hWire}
            \put(8,2){\hWire}
            \put(10,0){\boxGate{$R_z$}}
            \put(10,2){\boxGate{$R_z$}}
            \put(12,0){\boxGate{$R_z$}}
            \put(12,0){\topC}
        \end{picture}
    \end{center}
  We now merge the $R_z$ gates with the $c$ and $d$
  placeholders; there remain 13 parameters --- three each in the
  $a,b,c,d$ placeholders and one in the controlled-$R_z$ gate.
  This circuit fails to be universal.
  }

  In a different direction, one may ask whether one can do better
  by measuring in a different basis.

  \begin{proposition} \label{prop:meas2qlbounds:anybasis}
     Consider the $1+1+1+1$ measurement don't care, $M$, corresponding
     to a given fixed basis. Some two-qubit operators
     cannot be simulated up to $M$ by a circuit
     with a single {\tt CNOT}.
  \end{proposition}
  \proof{
  We concatenate the circuit in question with a placeholder for a
  diagonal operator.
    
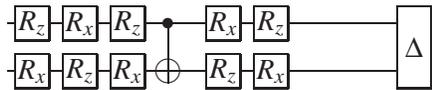
\begin{center}
        \begin{picture}(18,4)
            \put(0,0){\boxGate{$R_x$}}
            \put(0,2){\boxGate{$R_z$}}
            \put(2,0){\boxGate{$R_z$}}
            \put(2,2){\boxGate{$R_x$}}
            \put(4,0){\boxGate{$R_x$}}
            \put(4,2){\boxGate{$R_z$}}
            \put(6,0){\topCNOT}
            \put(8,0){\boxGate{$R_z$}}
            \put(8,2){\boxGate{$R_x$}}
            \put(10,0){\boxGate{$R_x$}}
            \put(10,2){\boxGate{$R_z$}}
            \put(12,0){\hWire}
            \put(12,2){\hWire}
            \put(14,0){\hWire}
            \put(14,2){\hWire}
            \put(16,0){\boxGateTwo{$\Delta$}}
        \end{picture}
    \end{center}
  Counting parameters gives 13 (the placeholder for the
  diagonal operator counts for three.) Thus this circuit cannot be
  universal.
  }

  Finally, we prove constructively that an arbitrary
  two-qubit operator can be implemented up to any
  CB-measurement don't care with a circuit
  containing two {\tt CNOT} gates and various one-qubit gates.

  \begin{proposition}
    \label{prop:univ}
    The $2$-qubit circuit given below is universal up to any
    CB-measurement don't-care.
    \noindent
    
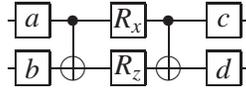
\begin{center}
    \begin{picture}(10,4)
    \put(0,0){\boxGate{b}}
    \put(0,2){\boxGate{a}}
    \put(2,0){\topCNOT}
    \put(4,0){\boxGate{$R_z$}}
    \put(4,2){\boxGate{$R_x$}}
    \put(6,0){\topCNOT}
    \put(8,0){\boxGate{d}}
    \put(8,2){\boxGate{c}}
    \end{picture}
    \end{center}
  \end{proposition}
    \proof{
    Consider a measurement with respect to any CB-subspace
    decomposition. The number and the probabilities of outcomes
    cannot change if we first measure along the $1+1+1+1$
    subspace decomposition. Indeed, the number of outcomes
    is determined by the number of subspaces in the last
    measurement, and the probabilities of outcomes for a given
    pure state by squared norms of projections onto those subspaces.
    In a CB-subspace decomposition, the squared norm of a projection
    onto a 2- or 3-dimensional subspace equals, by the Pythagorean theorem,
    the sum of squared norms of projections onto the computational-basis
    vectors in that subspace.  Therefore, a circuit which is
    universal up to a $1+1+1+1$ CB-measurement don't-care is
    universal up to any other CB-measurement don't-care, and it
    suffices to consider the $1+1+1+1$ case.

    Recall that the circuit of Fig. \ref{fig:18:cxz} is universal.
    As adding a reversible constant gate (e.g., {\tt CNOT}) to the end does not
    affect universality, the circuit below is universal as well.

    
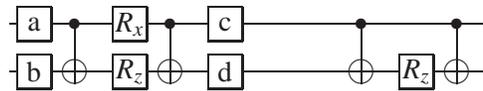
\begin{center}
    \begin{picture}(20,4)
    \put(0,0){\boxGate{b}}
    \put(0,2){\boxGate{a}}
    \put(2,0){\topCNOT}
    \put(4,0){\boxGate{$R_z$}}
    \put(4,2){\boxGate{$R_x$}}
    \put(6,0){\topCNOT}
    \put(8,0){\boxGate{d}}
    \put(8,2){\boxGate{c}}
    \put(10,0){\hWire}
    \put(10,2){\hWire}
    \put(12,0){\hWire}
    \put(12,2){\hWire}
    \put(14,0){\topCNOT}
    \put(16,0){\boxGate{$R_z$}}
    \put(16,2){\hWire}
    \put(18,0){\topCNOT}
    \end{picture}
    \end{center}

    Observe that the right portion of this circuit simulates a
    diagonal operator, which preserves the subspaces spanned by the
    computational basis vectors. Thus, by the discussion earlier in
    the section, the left portion of this circuit is universal up to
    measurement in the computational basis.
  }

  In applications such as Quantum Key Distribution, one may not know
  in advance which
  basis to measure in, but rather that one will choose at random
  between a given pair of bases for measurement. To save gates in
  this context, one could maintain two different circuits, one for
  each type of measurement. While it may seem counterintuitive
  that building two circuits would save on gates, note that the
  ``circuit'' here may consist of classical instructions to initiate a given
  laser pulse at a given time, thus we may maintain as many as we like
  in the memory of the classical computer we are using to control
  the quantum system. At issue is the execution time, which will be smaller
   when applying either of two smaller circuits (depending
   on the desired measurement) rather than a common,
   larger circuit followed by one of two measurements.

  An alternative approach to saving gates in such a context is
  to try and find circuits which simulate the desired operator up
  to either of the possible measurements. The only fact we used
  about the computational basis in the proof of Proposition
  \ref{prop:univ} was that operators expressible as $C_2^1 (I \otimes
  R_z(\theta)) C_2^1$ are diagonal in the computational basis.
  Such operators are also diagonal in any basis in which each
  vector lies in either span$(\ket{0},\ket{3})$ or
  span$(\ket{1},\ket{2})$. In particular, this includes bases of Bell
  states.

  \begin{proposition} Two {\tt CNOT}s suffice to simulate any
  two-qubit operator up to any measurement in a not necessarily
  predisclosed basis in which each vector lies in either span$(\ket{0},\ket{3})$ or
  span$(\ket{1},\ket{2})$.
  \end{proposition}

\section{Conclusions}
\label{sec:conclusions}

   Algorithms and lower bounds for quantum circuit synthesis
   have significantly advanced in the last two years.
   In particular, several universal two-qubit circuits with optimal
   gate counts are available
   \cite{Vidal:04,Vatan:04,Shende:18gates:04,Shende:cnotcount:04,
   Zhang:Bgate:04}, and, in the general
   case of $n$-qubit circuits, asymptotically optimal gate counts
   can be realized by matrix-decomposition algorithms
   \cite{Vartiainen:04, Mottonen:04}.

   In this context, we recall that quantum algorithms and
   cryptographic protocols often apply measurements, known in advance,
   after reversible quantum circuits. This allows a greater
   variety of circuits to be functionally equivalent, and
   we prove that useful information about measurement often
   facilitates finding smaller circuits. Taking into account
   a known input state also decreases circuit sizes.
   Both cases can be viewed as circuit synthesis
   for incompletely specified operators.

  Our work has parallels in synthesis of classical irreversible
  logic circuits, where truth tables are sometimes underspecified,
  and the synthesis program must complete them so as to allow for
  smaller circuits. In other words, outputs produced for some input
  combinations can
  be arbitrary. Such unspecified behaviors of classical circuits are
  traditionally called ``don't-cares''.
  While covered in undergraduate circuits
  courses, they remain a worthy subject of research and
  appear in a variety of circumstances in practice. For example,
  if a given circuit operates on outputs of another circuit,
  the latter may not be able to produce certain combinations of bits.
  While this cannot happen with reversible quantum circuits,
  we may nonetheless know in advance that
  the input state will be $\ket{0}$. Indeed, it may happen
  that the purpose of the circuit all along was to prepare a given
  state form $\ket{0}$. To this end, we point out that an $n$-qubit
  state can be prepared using $O(2^n)$ gates --- which is
  asymptotically optimal --- whereas $O(4^n)$ gates are necessary to
  simulate a generic $n$-qubit unitary operator. We also show that
  at most one maximally entangling gate is necessary and sufficient
  to prepare a $2$-qubit state, and, in particular, that a
  single {\tt CNOT} suffices. We have also shown that, if the final
  measurement is known to be in the computational basis, only two
  {\tt CNOT} gates are necessary.

\section*{Acknowledgements}
 \noindent

   This work is funded by the DARPA QuIST program and an NSF grant.
   The views and conclusions contained herein are those of the authors
   and should not be interpreted as neces\-sarily representing official
   policies or endorsements of employers and funding agencies.

\singlespace

\end{document}